\documentclass[modern]{aastex63}
\usepackage{epstopdf}
\usepackage{graphicx}
\usepackage{subfigure}
\usepackage{chngcntr}
\usepackage{lipsum}
\usepackage{float}
\usepackage{tabularx}
\usepackage{mathtools}
\usepackage{verbatim}
\usepackage{lineno}

\usepackage[nameinlink]{cleveref}

\newcommand{\y}[1]{{\textcolor{red}{#1}}}

\received{\today}
\revised{\today}
\accepted{\today}
\submitjournal{ApJ}
\shortauthors{Liang, Yoon and Zhao}

\begin{document}

\title{Moving Groups in the Solar Neighborhood with Gaia, APOGEE, GALAH, and LAMOST: Dynamical Effects Gather Gas and the Ensuing Star Formation Plays an Important Role in Shaping the Stellar Velocity Distributions}

\author[0000-0001-9283-8334]{Xilong Liang}
\affiliation{Yunnan Observatories, Chinese Academy of Sciences, 396 Yangfangwang, Guandu District, Kunming, 650216, People\'s Republic of China}
\affiliation{Department of Astronomy \& Center for Galaxy Evolution Research, Yonsei University, Seoul 03722, Republic of Korea}
\author[0000-0002-1842-4325]{Suk-Jin Yoon}
\affiliation{Department of Astronomy \& Center for Galaxy Evolution Research, Yonsei University, Seoul 03722, Republic of Korea}
\author[0000-0003-2868-8276]{Jingkun Zhao}
\affiliation{Key Laboratory of Optical Astronomy, National Astronomical Observatories, Chinese Academy of Sciences, Beijing 100012, China}

\correspondingauthor{Suk-Jin Yoon}
\email{sjyoon0691@yonsei.ac.kr}

\begin{abstract}

With Gaia, APOGEE, GALAH, and LAMOST data, we investigate the positional, kinematic, chemical, and age properties of nine moving groups in the solar neighborhood.
We find that each moving group has a distinct distribution in the velocity space in terms of its metallicity, $\alpha$ abundance, and age.
Comparison of the moving groups with their underlying background stars suggests that they have experienced the enhanced, prolonged star formation.
We infer that any dynamical effects that gathered stars as a moving group in the velocity space also worked for gas.
We propose for the first time that the ensuing newborn stars from such gas inherited the kinematic feature from the gas, shaping the current stellar velocity distributions of the groups.
Our findings improve the understanding of the origins and evolutionary histories of moving groups in the solar neighborhood.

\end{abstract}

%%https://github.com/astrothesaurus/UAT/blob/master/UAT.csv

\section{INTRODUCTION}\label{sec1}

Moving groups in the solar neighborhood provide valuable insights into the formation and evolution of our Galaxy's disk \citep{1996AJ....112.1595E,1998AJ....115.2384D}, thanks to their retention of birth environment signatures and dynamical histories. Moving groups are associated with large kinematic structures like arches, ridges, and phase spirals \citep{2018Natur.561..360A,2018A&A...616A..11G,2018A&A...619A..72R,2019MNRAS.488.3324F,2019MNRAS.489.4962K,2019A&A...626A..41M,2020A&A...643L...3L,2022A&A...667A.116B,2022MNRAS.516L...7H,2022A&A...663A..38K}.
Models have shown that many mechanisms affect the velocity distribution of stars in the solar neighborhood, such as (a) spiral waves churning both stars and gas, inducing radial mixing \citep{2002MNRAS.336..785S,2009MNRAS.396..203S}; (b) the effect of spiral arms \citep{2005AJ....130..576Q,2009ApJ...700L..78A,2011Natur.477..301P,2018MNRAS.481.3794H,2018ApJ...863L..37M,2020ApJ...888...75B}; (c) nonlinear effects of orbital resonances \citep{1986A&A...155...11C,2001A&A...373..511F,2009MNRAS.396L..56M,2019A&A...626A..41M,2022MNRAS.512.2171C,2024A&A...686A..92B} like the outer Lindblad resonance of the Galactic bar \citep{2000AJ....119..800D,2018MNRAS.477.3945H,2019MNRAS.488.3324F,2019MNRAS.482.1983F,2019A&A...626A..41M,2022MNRAS.514..460A,2022MNRAS.509..844T}, corotation resonance of a long bar \citep{2017ApJ...840L...2P,2017MNRAS.466L.113M,2019A&A...632A.107M,2020MNRAS.499.2416A,2020ApJ...890..117D}, and resonance sweeping by a decelerating Galactic bar \citep{2021MNRAS.500.4710C,2021MNRAS.505.2412C}; (d) heating of Galactic disks by mergers \citep{1993ApJ...403...74Q}; (e) footprints of the Sagittarius dwarf galaxy passing through the disk \citep{1993ApJ...403...74Q,2011Natur.477..301P,2016ApJ...823....4D,2019MNRAS.485.3134L,2021MNRAS.505.2561C}; (f) phase-mixing due to external perturbations \citep{2009MNRAS.396L..56M,2012MNRAS.419.2163G,2015MNRAS.454..933D,2018Natur.561..360A,2018MNRAS.481.3794H,2018A&A...619A..72R,2019MNRAS.485.3134L,2020A&A...643L...3L,2019MNRAS.489.4962K}; and (g) tidally stripped stars \citep{1999Natur.402...53H,2013MNRAS.433.1813S}.

The chemical tagging technique \citep{2002ARA&A..40..487F} is a popular method to uncover the fossil evidence provided by the Galaxy.
\citet{2007ApJ...655L..89B} studied the Hercules moving group with high-resolution spectra of nearby F and G dwarf stars, supporting its dynamical origin. \citet{2007AJ....133..694D} studied the age and chemical abundances of the HR 1614 group and suggested that it is the remnant of a dispersed star-forming event. \citet{2011MNRAS.415..563D} used high-resolution spectra to study the Hyades moving group and found that it is made up at least partly of dispersed Hyades cluster stars. \citet{2017ApJ...844..152L} utilized data from LAMOST DR3 combined with Gaia DR1 TGAS to unveil a distinct metal-poor gap separating the Hyades-Pleiades and Hercules moving groups, hinting at underlying Galactic processes shaping their compositions.
\citet{2019MNRAS.484..125R} studied K giants of the Hyades and Sirius superclusters and suggested that the Hyades moving group is populated by stars from the Hyades and Praesepe open clusters, while the Sirius moving group is populated by stars from the Ursa Major open cluster.
\citet{2021ApJ...922..105Y} investigated the chemistry and ages of moving groups using APOGEE DR16, revealing significant dispersion among them.
\citet{2023ApJ...956..146L} suggested that the Hercules 1 and Hercules 2 moving groups have formed primarily through radial migration processes according to their radial age and chemical abundance distributions.
\citet{2024A&A...681L...8N} proposed a theory that the bar could affect the star formation in the solar neighborhood.
\citet{2022A&A...663A..38K} discovered some moving groups exhibiting slightly enriched mean metallicity and periodic variations correlated with angular momentum.
They further suggested that these patterns in the ($R$, $V_\phi$) plane \citep{2019MNRAS.489.4962K,2019ApJ...887..193L,2020MNRAS.494.5936F} are linked to the large-scale Milky Way spiral arms and main bar resonances.
\citet{2023AJ....165..110Y} suggested that bar dynamics are not enough to explain all ridge properties in the ($R$, $V_\phi$) plane.
\citet{2022ApJ...935...28W} explored the chemical signatures of bar resonances and associated some moving groups with the bar's outer Lindblad and corotation resonances.
\citet{2023A&A...671A..56K} using chemodynamical simulations show that dynamical effects play a key role in the formation of large-scale metallicity variations across spiral arms.
\citet{2024A&A...690A.147H} used simulations to show that bar formation in galaxies is accompanied by an episode of radial migration induced by the bar's slowing down. This process is responsible for spreading stars from the inner disc toward the outer Lindblad resonances.
Despite the efforts, the mechanisms governing the formation and evolution of moving groups, as well as their connections to broader Galactic structures, remain unresolved. To address the issue, we undertake a comprehensive analysis of the chemical abundances, ages, and kinematic properties of moving groups within the thin disk, aiming to unravel their complex origins and evolution histories.

In this study, we adopt the terminology of moving groups in \citet{2017ApJ...844..152L}.
Some moving groups listed in \citet{2017ApJ...844..152L} are not studied in this work because they have only a small number of member stars with available chemical abundances and age information.
We combine the Hyades moving group and the Pleiades moving group as one entity due to their analogous characteristics and the precision of available data. \citet{2023MNRAS.519..432L} suggested that the Pleiades moving group's radial extent is only 200 pc, influenced by contributions from the Pleiades cluster's tidal tail \citep{2019A&A...628A..66L}.
Similarly, we do not distinguish between the two density peaks within the $\gamma$ Leo moving group. \citet{2018ApJ...863....4L} used high-resolution spectra observed with the Subaru High Dispersion Spectrograph to study the chemical abundances of the $\gamma$ Leo moving group and found that it is composed of local stars.
\citet{2018RNAAS...2...32M} suggested that the Coma Berenices moving group must have formed less than 1.5 billion years ago, possibly related to a pericentric passage of the Sagittarius dwarf satellite galaxy.
\citet{2021A&A...647A..19T} showed that the Pleiades, Hyades, and Coma Berenices moving groups are more populated by young clusters, while the Sirius region seems to contain a clump of clusters with ages exceeding 300 Myr, and no cluster populates in the two Hercules moving groups.
\citet{2023ApJ...956..146L} showed that Hercules 1 and Hercules 2 moving groups are two independent structures with different metallicities and ages.
\citet{2024A&A...686A..92B} measured the azimuthal and radial gradients of the moving groups and compared them with simulation results. They suggested that a simple model with a fast and a slow bar is not sufficient to produce the radial gradient of the Hercules moving groups, although it can produce the azimuthal gradient.

In Section \ref{sec2}, we describe the data used for selecting member stars and underlying background stars for each moving group, as well as the data used for determining their metallicity, $\alpha$ abundance, and age.
Section \ref{sec3} presents [Fe/H], [$\alpha$/Fe], and age distributions in the ($V_R$, $V_{\phi}$) coordinate. We also provide the comparison results of the member stars and the background sample of each moving group.
In Section \ref{sec4}, we discuss various theories in the literature that are relevant to our findings.
Finally, Section \ref{sec5} summarizes our new findings and their implications.

%\clearpage

\begin{deluxetable*}{lcccccl}
\tablecaption{The Properties of the Moving Groups Examined in This Study.\label{t1}}
\tablewidth{0pt}
\tablehead{
\colhead{Name} & \colhead{Mean } & \colhead{Mean } & \colhead{Mean } & \colhead{Number of } & \colhead{Number of }& \colhead{Colors in } \\
\colhead{} & \colhead{[Fe/H]} &\colhead{[$\alpha$/Fe]} &\colhead{age} &\colhead{Mem. stars}& \colhead{Bkg. stars}& \colhead{Plots}\\
\colhead{} & \colhead{(dex)} &\colhead{(dex)} &\colhead{(Gyr)} &\colhead{}& \colhead{}& \colhead{}
}
\decimalcolnumbers
\startdata
Hyades-Pleiades   &  0.018 $\pm$0.200 & -0.0087 $\pm$0.044 & 4.83 $\pm$2.67 & 448\,593& 399\,343& black  \\
Sirius            & -0.061 $\pm$0.186 & -0.0056 $\pm$0.044 & 4.57 $\pm$2.62 & 210\,019& 150\,636& magenta \\
Hercules 2        & -0.007 $\pm$0.226 &  0.0028 $\pm$0.043 & 5.83 $\pm$2.77 & 133\,281& 193\,235& violet \\
Coma Berenices    & -0.034 $\pm$0.183 & -0.0094 $\pm$0.045 & 4.58 $\pm$2.66 & 130\,992& 213\,033& green \\
Hercules 1        & -0.024 $\pm$0.211 &  0.0033 $\pm$0.045 & 5.57 $\pm$2.77 & 62\,680 & 89\,070& pink \\
Wolf 630          &  0.006 $\pm$0.219 & -0.0022 $\pm$0.043 & 5.36 $\pm$2.68 & 56\,579 & 43\,333& blue \\
$\gamma$ Leo      & -0.092 $\pm$0.214 &  0.0105 $\pm$0.044 & 5.85 $\pm$2.69 & 38\,824 & 88\,236& cyan \\
Dehnen98-6        & -0.038 $\pm$0.210 &  0.0025 $\pm$0.041 & 5.59 $\pm$2.69 & 33\,614 & 64\,940& red \\
Antoja12-GCSIII-13& -0.186 $\pm$0.223 &  0.0231 $\pm$0.048 & 6.53 $\pm$2.72 & 10\,098 & 16\,655& yellow \\
\enddata

\tablecomments{Columns (2), (3), and (4) are respectively, the mean metallicity, mean $\alpha$ abundance, and mean age of member stars in each moving group, while the $\pm$ values are the standard deviations. Columns (5) and (6) are, respectively, the total number of member stars and the number of background stars for each moving group. The last column (7) lists the colors of ellipses used in Figures \labelcref{mem,feh,afe,age,oc} for the corresponding moving groups.}
\end{deluxetable*}

\section{DATA} \label{sec2}

\subsection{Positions and Velocities}

\begin{figure*}
\includegraphics[width=1.1\textwidth, angle=0]{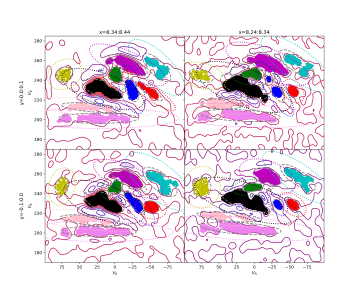}
\caption{
Velocity distributions of stars in spatial bins. We show here local four panels as an example; all panels for our total sample are shown in Figure \ref{amem}. The $x$ and $y$ spatial ranges in kiloparsec covered by each panel are denoted at the top and left of the figure. The contour represents wavelet transformation coefficients, and the colored dots represent selected member stars of each moving group. The concentric ellipses show the velocity range used to select the background sample for each moving group. The color used for each moving group is listed in Table \ref{t1}, column (7).
}
\label{mem}
\end{figure*}

We select stars from the Gaia DR3 catalog \citep{2016A&A...595A...1G, 2023A&A...674A...1G} with $parallax\_over\_error \geq 5$ and $ruwe<1.4$, utilizing the astrometric data (parallax, proper motion, and radial velocity) to obtain positions and velocities with the Python package, Astropy \citep{2022ApJ...935..167A}.
The spatial coordinates, denoted by $x$, $y$, and $z$, indicate the star's position within the Galactic Cartesian coordinate system. $V_R$, $V_{\phi}$, and $V_z$ represent the velocity components of the star within the Galactic cylindrical coordinate system. The Sun's location is specified as $(x, y, z) = (8340, 0, 20)$ pc.
We select nearby stars with the criteria of $7840 < x < 8840$ pc, $\mid y \mid < 500$ pc, $\mid z \mid < 200$ pc, and $\mid V_z \mid < 60 ~\textrm{km s}^{-1}$. The total sample of nearby stars is divided into 100 $\times$ 100 pc spatial bins in the ($x$, $y$) coordinate. Figure \ref{mem} shows the velocity distribution of stars in the spatial bins close to the Sun as examples. The $x$ and $y$ spatial ranges in kiloparsec covered by each panel are denoted at the top and left of the figure. Figure \ref{amem} in the Appendix displays all the spatial bins for the total sample.

For stars in each spatial bin, we adopt the wavelet transformation to reveal overdensities in stars' velocity distribution in the ($V_R$, $V_{\phi}$) coordinate. The contour in each panel of Figure \ref{mem} shows the distribution of wavelet transformation coefficients. The colored dots represent manually selected member stars for each moving group using the contour, with different colors representing different moving groups. This method of selection does not guarantee that all member stars are selected for each moving group, but we believe it is sufficient to statistically reveal chemical and age properties for each moving group.
The gray dashed ellipses are plotted using the Matplotlib package to approximately represent the velocity region covered by the member stars, using covariance confidence ellipses. The colored dotted ellipses are twice the size of the dashed ellipses. Stars between the concentric two ellipses are selected as a background sample for each corresponding moving group. We note that the member stars of the given moving group and of other moving groups have been removed from the background sample. The color used for each moving group is listed in Table \ref{t1}, column (7).

\subsection{Chemical Abundances and Ages}

The inclusion of chemical abundance and age information enables a comprehensive analysis of the moving group properties.
To obtain the working catalogs of stars with [Fe/H] and [$\alpha$/Fe], we cross-match\footnote{Different catalogs are cross-matched using the Topcat software \citep{tay05}.} three different catalogs: APOGEE DR17 \citep{2022ApJS..259...35A}, GALAH DR3 \citep{2021MNRAS.506..150B}, and LAMOST DR9 \citep{2012RAA....12.1197C, 2012RAA....12..723Z, 2015RAA....15.1095L}.
On the one hand, we construct a [Fe/H] catalog by combining the APOGEE, GALAH, and LAMOST data.
First, we construct the APOGEE/GALAH-combined [Fe/H] catalog.
To this, we identify 23,238 common thin disk stars on the [Fe/H] versus [$\alpha$/Fe] diagram based on APOGEE and GALAH catalogs.
To calibrate GALAH to APOGEE, we derive the difference between the mean values of [Fe/H] of the common stars from the two catalogs, i.e., $\textit{mean}([\rm{Fe/H}]_{\rm{GALAH}}) - \textit{mean}([\rm{Fe/H}]_{\rm{APOGEE}}) = -0.007$, and subtract the difference from [Fe/H] of all GALAH stars in a lump.
We combine the APOGEE [Fe/H] data and the calibrated GALAH [Fe/H] data, with duplicated sources removed.
Second, we construct the APOGEE/GALAH/LAMOST-combined [Fe/H] catalog.
To this, the LAMOST catalog is cross-matched with the APOGEE/GALAH combined catalog, obtaining 49,411 common stars.
To calibrate LAMOST to APOGEE, we derive the difference between the mean values of [Fe/H] of the common stars from the two catalogs, i.e., $\textit{mean}([\rm{Fe/H}]_{\rm{LAMOST}}) - \textit{mean}([\rm{Fe/H}]_{\rm{APOGEE}}) = 0.005$, and subtract the difference from [Fe/H] of all LAMOST stars in a lump.
We combine the APOGEE/GALAH [Fe/H] data and the calibrated LAMOST [Fe/H] data, with duplicated sources removed, to construct the final APOGEE/GALAH/LAMOST-combined [Fe/H] catalog.
On the other hand, we construct a [$\alpha$/Fe] catalog by combining the APOGEE and GALAH.
Note that LAMOST does not provide [$\alpha$/Fe].
For the aforementioned APOGEE--GALAH cross-matched 23,238 common thin disk stars, we derive the difference between the mean values of [$\alpha$/Fe] for common stars from the two catalogs, i.e., \textit{mean}$([\rm{\alpha/Fe}]_{\rm{GALAH}}) - \textit{mean}([\rm{\alpha/Fe}]_{\rm{APOGEE}}) = 0.011$, and subtract the difference from [$\alpha$/Fe] of all GALAH stars in a lump.
Then, we combine the APOGEE [$\alpha$/Fe] data and the calibrated GALAH [$\alpha$/Fe] data, with duplicated sources removed, to construct the final APOGEE/GALAH-combined [$\alpha$/Fe] catalog.

Age information is obtained from eight spectroscopic surveys \citep{2023A&A...673A.155Q} using the StarHorse code \citep{2018MNRAS.476.2556Q}.
The StarHorse provides stellar ages for approximately 2.5 million main-sequence turnoff and subgiant-branch stars, with typical age uncertainties around $30\%$.
The StarHorse is based on Bayesian inference, computing the marginal posterior distributions for the data given a set of stellar models and input stellar parameters. Validation against open cluster ages and other attempts to derive isochrone ages demonstrate that their ages are reliable for stars older than 2 Gyr.
We combine all catalogs from the StarHorse with duplicated sources removed.

In summary, we are left with 450\,247, 110\,008 and 328\,347 stars for the [Fe/H], [$\alpha$/Fe] and age samples, respectively.

\section{RESULTS} \label{sec3}
\subsection{Properties of the Moving Groups}

We provide the properties of the moving groups in Table \ref{t1}, including their mean metallicity, mean $\alpha$ abundance, and mean age of member stars, along with the corresponding standard deviations. The standard deviations represent internal dispersion within each moving group. Additionally, the table includes the total number of member stars and the number of background stars available for each moving group.

Consistent with the findings of \citet{2017ApJ...844..152L}, the Hyades-Pleiades and Wolf 630 moving groups exhibit the highest mean metallicity, while the Antoja12-GCSIII-13 moving group shows the lowest metallicity. Notably, the $\gamma$ Leo and Sirius moving groups also display relatively low metallicity. The Hercules 2 and Antoja12-GCSIII-13 moving groups show the widest dispersions in metallicity, whereas the Sirius and Coma Berenices moving groups exhibit the smallest metallicity dispersions.
The order of mean $\alpha$ abundance across moving groups does not directly mirror that of their metallicities.
The Antoja12-GCSIII-13 and $\gamma$ Leo moving groups have the richest $\alpha$ abundances, while the Coma Berenices and Hyades-Pleiades moving groups have the poorest $\alpha$ abundances.
The Antoja12-GCSIII-13 moving group has the highest dispersion in $\alpha$ abundance, while the Dehnen98-6 moving group has the smallest $\alpha$ abundance dispersion. Other moving groups demonstrate similar levels of $\alpha$ abundance dispersions.

The ages of moving groups do not adhere to a strict inverse correlation with their metallicities or a direct correspondence with their $\alpha$ abundances.
Notably, the Antoja12-GCSIII-13 moving group has the oldest mean age, aligning consistently with its ranking in terms of the aforementioned parameters.
The Sirius, Coma Berenices, and Hyades-Pleiades moving groups are relatively young, and take up the main part in the ($V_R$, $V_{\phi}$) coordinate. The Hercules 2 and Hercules 1 moving groups have the largest age dispersions, while the Sirius moving group has the smallest age dispersion.
Overall, the properties of the moving groups (Table \ref{t1}) suggest that it is important to consider multiple stellar parameters when investigating the origins and evolution history of moving groups in the Milky Way.

\subsection{The Moving Groups with Respect to [Fe/H], [$\alpha$/Fe], and Ages in the Velocity Space}

 \begin{figure*}
 \hspace*{-0.5in}
 \includegraphics[width=1.1\textwidth, angle=0]{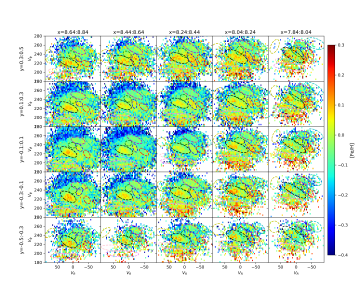}
 \caption{
 The metallicity distribution in the ($V_R$, $V_\phi$) coordinate. The $x$ and $y$ spatial ranges in kiloparsec covered by each panel are denoted at the top and left. The color bar on the right indicates the color corresponding to [Fe/H] values for all panels. The ellipses approximately represent the velocity range covered by each moving group. The colors used for ellipses are listed in Table \ref{t1}, column (7).
 }
 \label{feh}
 \end{figure*}

 \begin{figure*}
 \hspace*{-0.5in}
 \includegraphics[width=1.1\textwidth, angle=0]{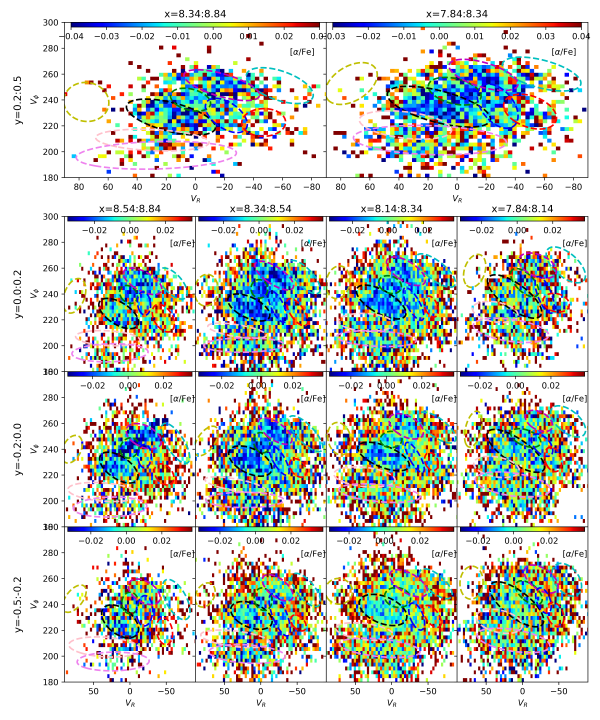}
 \caption{
 The same as Figure \ref{feh} but for $\alpha$ abundance. The color bar at the top of each panel indicates the color corresponding to the [$\alpha$/Fe] values.
 }
 \label{afe}
 \end{figure*}

 \begin{figure*}
 \hspace*{-0.5in}
 \includegraphics[width=1.2\textwidth, angle=0]{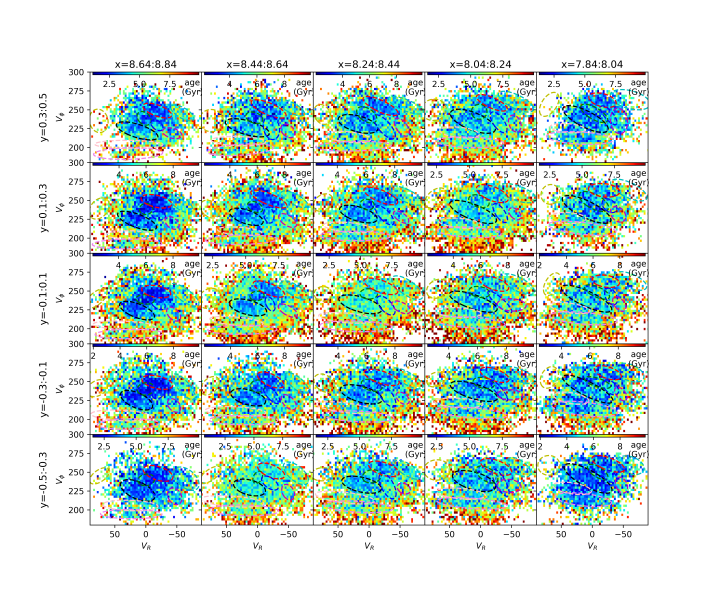}
 \caption{
 The same as Figure \ref{feh} but for age. The color bar at the top of each panel indicates the color corresponding to the age values.
 }
 \label{age}
 \end{figure*}

The metallicity, $\alpha$ abundance, and age samples are divided into spatial bins in the ($x$, $y$) coordinate. The $x$ and $y$ spatial ranges in kiloparsec covered by each panel are denoted at the top and left of Figures \labelcref{feh,afe,age}. Within each spatial bin, stars are further divided into 1 km s$^{-1}$ square velocity bins in the ($V_R$, $V_\phi$) coordinate. Figure \ref{feh} depicts the distributions of the metallicity sample in the ($V_R$, $V_\phi$) coordinate. The median [Fe/H] value is calculated to represent each velocity bin.  The color bar on the right indicates the color corresponding to [Fe/H] values for all panels. Velocity bins with fewer than two stars are left blank. The dashed ellipses are plotted using the same method as the dashed ellipses in Figure \ref{mem} to approximately represent the velocity regions covered by member stars of moving groups, using covariance confidence ellipses. The corresponding colors used for ellipses are listed in Table \ref{t1}, column (7).

The $\alpha$ abundance sample (Figure \ref{afe}) and age sample (Figure \ref{age}) are analyzed similarly to the metallicity sample (Figure \ref{feh}). In Figure \ref{afe}, larger spatial bin sizes have to be adopted to include enough stars in each spatial bin. The color bar at the top of each panel in Figures \labelcref{afe,age} indicates the color corresponding to [$\alpha$/Fe] and age values for that panel, respectively. Figures \labelcref{feh,afe,age} collectively illustrate that moving groups tend to be rich in metallicity, low in $\alpha$ abundance, and young in the velocity space.

\subsection{Comparison of the Moving Groups with Their Underlying Background Stars}

 \begin{figure*}
 \includegraphics[width=1.0\textwidth, angle=0]{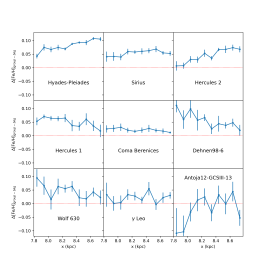}
 \caption{
 Metallicity comparison with the background sample. The title of each panel shows the name of the moving group.
 The dots with error bars represent mean values with uncertainties of $\Delta$[Fe/H] of different $y$ positions at a given $x$ position. The error bars are obtained through the jackknife analysis. $\Delta$[Fe/H] is defined as the median [Fe/H] of member stars minus the median [Fe/H] of the corresponding background star sample within an ($x$, $y$) spatial bin.
 }
 \label{dfeh}
 \end{figure*}

 \begin{figure*}
 \includegraphics[width=1.0\textwidth, angle=0]{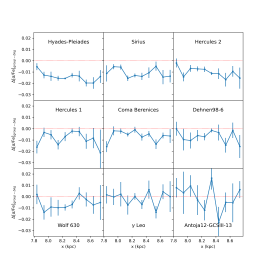}
 \caption{
 The same as Figure \ref{dfeh} but for $\alpha$ abundance.
 }
 \label{dafe}
 \end{figure*}

 \begin{figure*}
 \includegraphics[width=1.0\textwidth, angle=0]{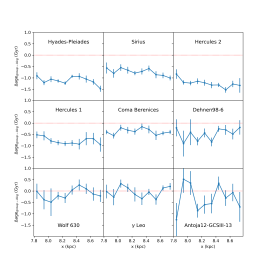}
 \caption{ The same as Figure \ref{dfeh} but for age.
 }
 \label{dage}
 \end{figure*}

To quantify the tendencies observed in Figures \labelcref{feh,afe,age}, background samples are utilized. Metallicity, $\alpha$ abundance, and age information are obtained by cross-matching member stars and background stars of each moving group with the metallicity sample, $\alpha$ abundance sample, and age sample.
In each panel of Figure \ref{amem}, $\Delta$[Fe/H] is calculated as the median [Fe/H] of member stars minus the median [Fe/H] of the corresponding background stars selected by the concentric ellipse. Consequently, we obtain 10 $\times$ 10 values of $\Delta$[Fe/H] for each moving group at different $x$-$y$ positions. Then, as illustrated in Figure \ref{dfeh}, the averaged value of $\Delta$[Fe/H] of different $y$ positions at a given $x$ position is plotted as a dot, and the error bar represents uncertainty obtained through the jackknife analysis.
The horizontal red dotted line in each panel indicates the position of $\Delta$[Fe/H] being zero. The same treatment is applied to the $\alpha$ abundance and age, and the results are shown in Figures \labelcref{dafe,dage}, respectively. We note that some moving groups do not have $\alpha$ abundance information at certain $x$-$y$ positions. We skip those spatial bins when calculating $\Delta$[$\alpha$/Fe].

As depicted in Figure \ref{dfeh}, most moving groups exhibit relatively richer metallicity than background stars.
Some $\Delta$[Fe/H] values for the $\gamma$ Leo moving group are reaching zero. The Antoja12-GCSIII-13 moving group shows, on average, no difference in metallicity distribution compared to its background sample.

Figure \ref{dafe} shows the distribution of the averaged values of $\Delta$[$\alpha$/Fe] of different $y$ positions at a given $x$ position, and the error bar represents uncertainty obtained through the jackknife analysis.
Most moving groups display, on average, relatively lower $\alpha$ abundance than background samples. The Hyades-Pleiades, Sirius, and Hercules 2 moving groups in the top row show significantly lower $\alpha$ abundances than their background samples.
At some $x$ positions, $\Delta$[$\alpha$/Fe] for the Hercules 1, Coma Berenices, and Dehnen98-6 is small.
The Wolf 630 moving group has positive $\Delta$[$\alpha$/Fe] values at two $x$ positions.
The Antoja12-GCSIII-13 and $\gamma$ Leo moving groups show, on average, no difference.

Figure \ref{dage} shows the distribution of the averaged values of $\Delta$age of different $y$ positions at a given $x$ position, and the error bar represents uncertainty obtained through the jackknife analysis.
The Hyades-Pleiades, Sirius, Hercules 2, Hercules 1, and Coma Berenices moving groups tend to have significantly younger ages than background samples.
The Dehnen98-6 moving group is also younger than its background sample, though $\Delta$age is reaching zero at certain $x$ positions.
The Wolf 630, $\gamma$ Leo, and Antoja12-GCSIII-13 moving groups in the bottom row do not exhibit significant differences from their background samples.

In summary, the mean $\Delta$[Fe/H] values for the top six moving groups in Figure \ref{dfeh} range from 0.02 to 0.08, mostly around 0.05. This means that more irons are produced by enhanced star formation compared to background stars. The mean $\Delta$[$\alpha$/Fe] values for the top six moving groups in Figure \ref{dafe} range from -0.006 to -0.014, mostly around -0.01. Meanwhile, the mean $\Delta$age values for the top six moving groups in Figure \ref{dage} range from -0.4 to -1.2 Gyr, mostly around -1 Gyr. The $\Delta$age range and $\Delta$[Fe/H] range agree reasonably well with the simulated age--metallicity relation for the solar neighborhood by \citet{1992BASI...20..177R}.

\section{DISCUSSION} \label{sec4}

The observed variations of metallicity, $\alpha$ abundance, and age among the moving groups in the velocity space distributions suggest diverse mechanisms influencing the formation and evolution of the groups.
The similarity between the $\gamma$ Leo and Antoja12-GCSIII-13 moving groups and their surrounding background stars in terms of the chemical and age distributions implies that these groups may have formed from locally born stars \citep{2018ApJ...863....4L}. These groups could have been dynamically gathered in the velocity space by gravitational interactions, such as the bar or spiral arms.
In contrast, the other moving groups exhibit distinct chemical and age signatures, suggesting more complex formation histories. We propose that these groups are not only affected by dynamical interactions but also by the accumulation of gas from the interstellar medium, leading to enhanced star formation. The richer metallicity, lower $\alpha$ abundance, and younger age support this hypothesis, indicating these groups have experienced more star formation episodes than their background stars.

Several mechanisms can increase the star formation rate in the solar neighborhood, including the passage through spiral arms \citep{1969ApJ...158..123R}, pericentric passage of the Sagittarius dwarf satellite galaxy \citep{2020NatAs...4..965R}, and long-term effects such as resonance sweeping by the decelerating bar \citep{2021MNRAS.505.2412C}, corotation resonance of the main spiral arms \citep{2017ApJ...843...48L}, and the outer Lindblad resonance of the bar \citep{1996ApJ...469..131E}.
Some simulations suggest that mergers like the Sagittarius dwarf galaxy can induce enhancement of star formation, but they may also dilute the metallicity of the central region, which is not evident from our observation \citep{2018MNRAS.479.3381B,2024MNRAS.527.2426A}.
Newborn stars via these mechanisms should not be randomly distributed in the velocity space but prefer regions taken by moving groups.
We assume that newborn stars triggered by transient density waves or shocks from the collision of a satellite galaxy with the disk carry distinct kinematic features compared to background stars.
Even after these stars fully disperse into the background, moving groups could inherit some of these features. We suggest that different moving groups may have been born from different historical events or may have inherited varying numbers of stars from those events.

\begin{figure*}
\includegraphics[width=1.0\textwidth, angle=0]{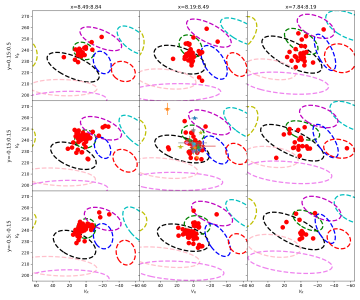}
\caption{
Comparison of velocity distributions of moving groups, open clusters, and young associations. Ellipses are moving groups. The colors used for ellipses are listed in Table \ref{t1}, column (7). Red circles are open clusters, while stars are young associations. The $x$ and $y$ spatial ranges in kiloparsec covered by each panel are denoted at the top and left.
}
\label{oc}
\end{figure*}

In Figure \ref{oc}, we show the velocity distributions of our moving groups along with open clusters and young associations. The colored ellipses represent the moving groups with corresponding colors listed in Table \ref{t1}, column (7).
Red circles are nearby open clusters listed in \citet{2023ApJS..265...12Q} with $ \mid z \mid < 200 $ pc.
Star symbols represent young associations within 150 pc from the Sun \citep{2018ApJ...856...23G}.
The $x$ and $y$ spatial ranges in kiloparsec covered by each panel are denoted at the top and left.
We confirm that open clusters and young associations tend to occupy the velocity regions associated with the Hyades-Pleiades (black ellipse) and Coma Berenices (green ellipse) moving groups. Additionally, some open clusters are found in the velocity regions of the Sirius (magenta ellipse), Dehnen98-6 (red ellipse), and Hercules 1 (pink ellipse) moving groups.
In the leftmost column, open clusters are predominantly found in the velocity region of the Coma Berenices moving group. Conversely, in the rightmost column, open clusters are more commonly associated with the Hyades-Pleiades moving group.
With the age information listed in \citet{2023ApJS..265...12Q}, we find that open clusters in the velocity region of the Coma Berenices moving group are, on average, younger than those in the velocity region of the Hyades-Pleiades and Sirius moving groups, which is consistent with our results (Table \ref{t1}, column (4)).
In the middle panel of the rightmost column, the open cluster in the velocity region of the Dehnen98-6 moving group is the oldest one in all open clusters plotted in Figure \ref{oc}, and the open cluster in the velocity region of the Hercules 1 moving group in the same panel is also old.
In this regard, \citet{2020AJ....159..196L} found a new moving group associated with the Orion star-forming complex. The mechanism behind such association is supposed to be density waves of the spiral arms.

Throughout the history of the solar neighborhood, many open clusters have been created during star formation events and subsequently dissolved due to natural evolution. Stars from dissolved open clusters and the tidal tails of open clusters may have not been completely digested by the Milky Way. Some stars can still share common kinematic features, explaining the observed connection between open clusters and some moving groups. This indicates that the past star formation events continue to influence the present-day velocity distributions of stars in the solar neighborhood.

\section{SUMMARY} \label{sec5}

This study examined the chemical abundances, ages, and kinematic properties of nine moving groups in the solar neighborhood.
Our investigation into the positional, kinematic, chemical, and age properties of the moving groups highlights the involvement of various mechanisms in their formation and evolution.

Through our investigation, several key findings have emerged:
\begin{itemize}
\item The observed metallicity, $\alpha$ abundance, and age of thin disk stars in the solar neighborhood have patterns that coincide with moving groups in the velocity coordinate.
\item The variations observed in metallicity, $\alpha$ abundance, and age among moving groups suggest diverse formation mechanisms.
\item The majority of our moving groups show chemical and age signatures distinct from surrounding background stars, which can be readily explained by enhanced star formation.
\item The mechanism behind the moving groups, whether it is radial migration processes, dynamical resonance of the bar and spiral arms, or pericentric passage of a satellite galaxy, should gather not only stars but also gas.
\item Dynamical effects gather gas and the ensuing star formation plays an important role in shaping the stellar velocity distributions.
\end{itemize}

By tracing the spatial and kinematic properties of moving groups, it is possible to constrain the processes driving stellar migration and star formation across different regions of the Galaxy.
Further investigations, including detailed chemical tagging studies and dynamical simulations, are needed to disentangle the relative importance of these mechanisms and understand their implications for the Galactic evolution.
We anticipate that the combined database from future data releases of Gaia, LAMOST, APOGEE, GALAH, and DESI \citep{2024MNRAS.533.1012K}, and upcoming surveys like CSST \citep{2021RAA....21...92S} and LSST \citep{2019ApJ...873..111I} will be helpful to better understand the Milky Way \citep{2024ApJ...970..121S}.

\acknowledgments
S.-J.Y. and X.L. acknowledge support from (1) the Mid-career Researcher Program (RS-2024-00344283) and (2) the Basic Science Research Program (2022R1A6A1A03053472) through the National Research Foundation of Korea.
J.Z. acknowledges support by (1) the National Natural Science Foundation of China (grant Nos.: 11988101, 12273055, and 11927804, (2) the global common challenge project of the Chinese Academy of Sciences (grant No.: 178GJHZ2022040GC), and (3) the support from the 2 m Chinese Space Station Telescope project.

This project is supported by the National Natural Science Foundation of China (NSFC, grant Nos. 12288102, 12125303, and 12090040/3), the National Key R\&D Program of China (grant No. 2021YFA1600403), the Yun-nan Fundamental Research Projects (grant No. 202201BC070003), the International Centre of Supernovae, Yunnan Key Laboratory (No. 202302AN360001), and Yunnan Revitalization Talent Support Program—Science \& Technology Champion Project (grant No. 202305AB350003).

This work has made use of data from the European Space Agency (ESA) mission
{Gaia} (\url{https://www.cosmos.esa.int/gaia}), processed by the {Gaia}
Data Processing and Analysis Consortium (DPAC,
\url{https://www.cosmos.esa.int/web/gaia/dpac/consortium}). Funding for the DPAC
has been provided by national institutions, in particular, the institutions
participating in the {Gaia} Multilateral Agreement.
%\email{aastex-help@aas.org}.

\appendix

Figure \ref{amem} is similar to Figure \ref{mem} but for our total sample.
Each subplot shows the $V_R$ versus $V_{\phi}$ distribution of stars in each spatial bin.
The $x$ and $y$ spatial ranges in kiloparsec covered by each panel are denoted at the top and left.
The position of the Sun is at $x=8.34$ kpc and $y=0$ kpc.
For stars in each bin, wavelet transformation is applied to stars' velocity distribution in the $V_R - V_{\phi}$ coordinate. The wavelet transformation coefficients are shown as contours in each panel. The colored dots represent manually selected member stars for each moving group, with different colors representing different moving groups. These concentric ellipses are used to select the background sample for each moving group.

%\end{document}

\counterwithin{figure}{section}

\renewcommand{\thefigure}{A\arabic{figure}}
\setcounter{figure}{0}

\begin{figure}
\centering
\includegraphics[width=1.1\textwidth, angle=0]{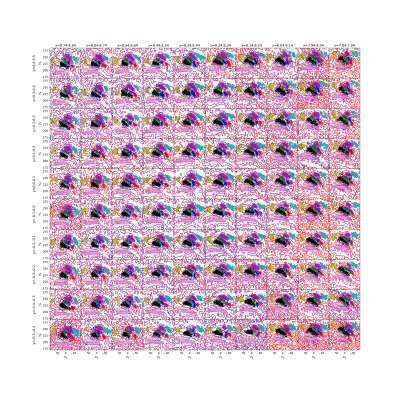}
\caption{Velocity distributions of stars in spatial bins for the total sample.
These subplots are similar to those in Figure \ref{mem}, but for all spatial bins. }
\label{amem}
\end{figure}

\end{document}